\documentclass[twocolumn,showpacs,preprintnumbers,amsmath,amssymb]{revtex4}

\usepackage{graphicx}
\usepackage[colorlinks,dvipdfm]{hyperref}

\usepackage{subfigure}

\usepackage{dcolumn}
\usepackage{bm}
\usepackage{amsthm}
\usepackage{amsmath}
\usepackage{extarrows}
\usepackage{multirow}
\usepackage{longtable}
\usepackage{rotating}

\begin{document}

\preprint{}

\title{Szegedy Quantum Walks with Memory on Regular Graphs}

\author{Dan Li$^{1,2,\ast}$, Ying Liu$^{1}$, Yu-Guang Yang$^{3}$, Juan Xu$^{1}$, Jia-Bing Yuan$^{1}$}
\affiliation{ $^{1}$  College of Computer Science and Technology, Nanjing University of Aeronautics and Astronautics, Nanjing, China\\
$^{2}$  Collaborative Innovation Center of Novel Software Technology and Industrialization, Nanjing, China\\
        $^{3}$ Faculty of Information Technology, Beijing University of Technology, Beijing 100124, China}

\date{\today}

\begin{abstract}

Quantum walks with memory(QWM) are a type of modified quantum walks that record the walker's latest path. The general model of coined QWM is presented in Phys. Rev. A 93, 042323 (2016). In this paper, we present general model of Szegedy QWM. Importantly, the relation of coined QWM and Szegedy QWM is revealed. By transforming coined QWM to Szegedy QWM, some amazing results about QWM are founded.
\end{abstract}

%
\maketitle

\section{Introduction}
\label{sec:level1}

Quantum walk is an essential model to realize quantum computation. Quantum walk provides a method to explore all possible paths in a parallel way due to constructive quantum interference along the paths. Many kinds of quantum walks  have been proposed, such as single-particle quantum walks \cite{101,102,103,104}, two-particle quantum walks \cite{105,106,107}, three-state quantum walks \cite{108,109}, controlled interacting quantum walks  \cite{110,111}, indistinguishable particle quantum walks \cite{112,113}, disordered quantum walk \cite{114,115},  etc. Each type of quantum walks has its own special features and advantages. Therefore, algorithms based on quantum walks have been established as a dominant technique in quantum computation, ranging from element distinctness \cite{117} to database searching \cite{118,119,120,121}, from constructing quantum hash functions \cite{110,111} to graph isomorphism testing \cite{122,123}.    

Most quantum walks been studied are quantum walks without memory(QWoM) on regular graphs. Quantum walks with memory(QWM)  have been studied in Ref.\cite{201,202,203,204,205}, while classical walks with memory have been used in research on the behavior of hunting, searching and building human memory search model. Standard QWM are a kind of modified quantum walks that have many extra coins to record the walker's latest path. Rohde et. al. presented a kind of coined QWM provided by recycled coins and a memory of the coin-flip history \cite{201}. Mc Gettrick presented another kind of coined QWM whose coin state decides the shift is `Reflect' or `Transmit' \cite{202,203}. Konno and Machida provided limit theorems for Mc Gettrick's QWM\cite{204}. Li and Mc Gettrick et. al. \cite{205} presented the generic model of coined quantum walks with memory by introducing the relation of QWM on a regular graph and QWoM on the line digraph of the regular graph. With this model, it becomes possible to build any wanted coined QWM on regular graphs and to study  properties of different kinds of QWM.

Except coined quantum walk, Szegedy quantum walk is also a very important version of quantum walks. Szegedy quantum walk was introduced by Szegedy by means of quantizing Markov chains described by some transition matrix \cite{301}. Later Portugal \cite{302} showed the connection between the coined quantum walk  and Szegedy's quantum walk. Konno \cite{303} introduced the notion of partition-based quantum walks, and he proved that the two-step coined quantum walks, an extension of Szegedy quantum walks for multigraphs, and the two-tessellate staggered quantum walks are unitary equivalent. Liu et. al. \cite{207} constructed Szegedy quantum walks  on regular uniform hypergraphs.

It is proven that Szegedy quantum walk is used in a variety of different applications, such as verifying matrix products \cite{401}, searching triangles \cite{402},testing group commutativity \cite{403}, approximating the effective resistance in electrical networks \cite{404}, and quantum Pagerank algorithm \cite{405} for determining the relative importance of nodes in a graph.

In this paper, the generic model of Szegedy QWM is presented in Sect.\ref{sec:level2}. Then, the relation of Szegedy QWM and coined QWM is revealed  in Sect.\ref{sec:level3}. Importantly,  Szegedy QWM provides a useful tool in analyzing coined QWM. In the process of analyzing coined QWM with Szegedy QWM, we get some amazing facts in  Sect.\ref{sec:level4}. These facts subvert the existing cognition of coined QWM. Finally, a  conclusion is given in Sect.\ref{sec:level5}.

\section{Generic model of Szegedy QWM}
\label{sec:level2}

In this section, we present the general model of Szegedy QWM.

Here, some preparation are given. Let $G=(V,E)$ be a digraph with vertex set $V(G)$ and arc set $E(G)$.  With fixed labeling of vertices, the adjacency matrix of a digraph $G$ with $N$ vertices, denoted by $M(G)$, is the $N\times N$ (0,1)-matrix with \textit{ij}-th element defined by $M_{i,j}(G)=1$ if $(x_i,x_j)\in E(G)$ and $M_{i,j}(G)=0$, otherwise. The line digraph of a digraph $G$, denoted by $\overrightarrow{L}G$, is defined as follows: the vertex set of $\overrightarrow{L}G$ is $E(G)$; for $x_a$, $x_b$, $x_c$, $x_d \in V(G)$, $((x_a,x_b),(x_c,x_d))\in E(\overrightarrow{L}G)$ if and only if $(x_a,x_b)$ and $(x_c,x_d)$ are both in $E(G)$ and $x_b=x_c$. The line digraph of $\overrightarrow{L}G$ is denoted by $\overrightarrow{L}^2G$. Similarly, there are $\overrightarrow{L}^dG$s with $d\in N^\ast$. For simplicity,  all of them are called line digraph of $G$.

According to Ref.\cite{205}, coined quantum walks with $d$ memory on a regular graph $G$ can be transformed to coined QWoM on $\overrightarrow{L}^dG$. Similarly, Szegedy quantum walks with $d$ memory on a regular graph $G$ could be seen as a Szegedy QWoM on  $\overrightarrow{L}^dG$. Then the general model of Szegedy QWM is given as follows.

\textbf{Definition 1} Let $G$  an $m$-regular graph. Define $\pi$ be an \emph{edge dicycle partition} of $\overrightarrow{L}^dG$ such that
\begin{equation}\label{Eq201}
    \pi: \overrightarrow{L}^dG\rightarrow \{ C_1, C_2, \cdots\},
\end{equation}
where $\{C_k|k=1,\cdots\}$ satisfy that $\bigcap_k E(C_k)=\varnothing$, $\bigcup_kE(C_k)=E(\overrightarrow{L}^dG)$  and each $C_k$ is an eulerian digraph.  We denote the set of edge dicycle partitions of $\overrightarrow{L}^dG$ by $\Pi_{\overrightarrow{L}^dG}$.

\textbf{Definition 2} For $\pi\in\Pi_{\overrightarrow{L}^dG}$ with $\overrightarrow{L}^dG \stackrel{\pi}{\longrightarrow} \{ C_1, C_2, \cdots\}$, define $f$ such that for any $(v_1,v_2)\in E(\overrightarrow{L}^dG)$,
\begin{equation}\label{Eq202}
    f: (v_1,v_2)\rightarrow (v_2,v_3)
\end{equation}where $(v_1,v_2)$ and $(v_2,v_3)$ belong to same $E(C_k)$.

$\pi$ is an edge dicycle partition, hence $f$ is a bijection. $f$ requires the walker to walk along a subgraph  $C_k$.

\textbf{Definition 3} For a Szegedy QWoM  on the line digraph of $G$ denoted by $\overrightarrow{L}^{d}G$, i.e. Szegedy QWM on $G$, the evolution is decomposed into two steps, $U=S*R$, defined as
{\setlength\arraycolsep{2pt}
\begin{eqnarray} \label{Eq203}
     &&R=2\sum_v|\psi_v\rangle\langle \psi_v|-I \\
     &&S=\Sigma_{(v,w)\in E(\overrightarrow{L}^{d}G)}|f(v,w)\rangle\langle v,w|,\label{Eq204}
\hspace{1mm}
\end{eqnarray}}where \begin{equation}\label{Eq205}
    |\psi_v\rangle=|v\rangle\otimes \sum_w\sqrt{q_{vw}}|w\rangle.
\end{equation}$q_{vw}$ is the probability of leaping from $v$ to $w$.

Definition 3 presents the  Szegedy QW with $d$ memory on $G$. It is not only the new form of QWM, but also  a useful tool in analyzing coined QWM.

\section{Coined QWM and its relation with Szegedy QWM}
\label{sec:level3}

\subsection{Coined QWM}
\label{sec:level31}

To uncover the relation of Szegedy QWM and coined QWM, the model of coined QWM is introduced in Definition 4, 5, 6. More details please refer to Ref. \cite{205}.

\textbf{Definition 4} Let $G$  an $m$-regular graph. Define $\pi'$ be a \emph{partition} of $\overrightarrow{L}^dG$ such that
\begin{equation}\label{Eq301}
    \pi': \overrightarrow{L}^dG\rightarrow \{ C'_1, C'_2, \cdots, C'_m\},
\end{equation}
where $\{C'_k|k=1,\cdots,m\}$ satisfy that $V(C'_k)=V(\overrightarrow{L}^dG)$, $\bigcup_kE(C'_k)=E(\overrightarrow{L}^dG)$  and for every vertex $v\in V(C'_k)$, the outdegree is $1$. \emph{Dicycle partition} is a kind of partition which satisfies that for every vertex $v\in V(C_k)$, the outdegree and indegree are 1. The set of  partitions of $\overrightarrow{L}^dG$ are denoted by $\Pi'_{\overrightarrow{L}^dG}$.

\textbf{Definition 5} For $\pi'\in\Pi'_{\overrightarrow{L}^dG}$ with $\overrightarrow{L}^dG \stackrel{\pi'}{\longrightarrow} \{ C'_1, C'_2, \cdots, C'_m\}$, define
\begin{equation}\label{Eq302}
    f'_{C'_k}: V(\overrightarrow{L}^dG) \rightarrow V(\overrightarrow{L}^dG)
\end{equation}
such that for any $v\in V(\overrightarrow{L}^dG)$,
\begin{equation}\label{Eq303}
    (v,f'_{C'_k}(v))\in E(C'_k)
\end{equation}

\textbf{Definition 6} For a coined QWoM  on the line digraph of $G$ denoted by $\overrightarrow{L}^{d}G$, i.e. coined QWM on $G$, the evolution is decomposed into two steps, $U=D*C$, defined as
{\setlength\arraycolsep{2pt}
\begin{eqnarray} \label{Eq304}
     C: &&|v,c\rangle \longrightarrow \sum_j A_{c,c_j} |v,c_j\rangle; \\
     D: &&|v,c_j\rangle \longrightarrow |f'_{C'_j}(v),gc(v,c_j)\rangle \label{Eq305}
\hspace{1mm}
\end{eqnarray}}The coin shift function $gc$ has to satisfy {\setlength\arraycolsep{2pt}
\begin{eqnarray} \label{Eq306}
\{\overbrace{gc(v_i,c_k),\cdots}^m\}=\{c_1,c_2,\cdots,c_m\}\  with \ v=f'_{C_k'}(v_i).
\hspace{1mm}
\end{eqnarray}}

For a coined QWM, whether the  partition is a dicycle partition affects the choice of coin shift function. Through the analysis in \ref{sec:level32},  the gap between \emph{dicycle partition} and \emph{partition} is bridged by Szegedy QWM.

\subsection{The relation of coined QWM and Szegedy QWM}
\label{sec:level32}

Szegedy quantum walk is a kind of quantum walk that the walker wander on edges of the graph, while coined quantum walk is a kind of quantum walk that the walker wander on vertices of the graph. Szegedy QWM lives in the Hilbert space $H_E$ spanned by $|v,w\rangle$, where $(v,w)\in E(\overrightarrow{L}^{d}G)$. Coined QWM lives in the Hilbert space $H_{V,C}$ spanned by $|v,c\rangle$, where $v\in V(\overrightarrow{L}^{d}G)$, $c$ denotes the state of  coin. The movement of Szegedy QWM is controlled by the adjacent matrix, an \emph{edge  dicycle partition} of the graph, while  the movement of coined QWM is controlled by  a coin operator, a \emph{partition} of the graph and a coin shift function.

From the above mentioned, Szegedy QWM and coined QWM don't look the same. However, the relation of Szegedy QWM and coined QWM can be revealed by following analysis.

A brief summation is given in Table.\ref{Table1}.
\begin{table}[htbp]
\centering
\caption{Correspondence}\label{Table1}
\begin{tabular}{|c|c|}
\hline
  \ \ \ \ \ \ $H_E\ \ \ \ \ \ $  &\ \ \ \ \ \  $H_{V,C}$ \ \ \  \ \ \ \\
\hline
  $|v,w\rangle$ & $|v,c\rangle$  \\
\hline
  $S$ & $D$  \\
\hline
  $R$ & $C$  \\
\hline
\end{tabular}
\end{table}

1. For an $m$-regular graph $G$, $H_E$ and $H_{V,C}$ are the same size, $size(V(\overrightarrow{L}^{d}G))\cdot m$.

2. Let $v$ in $|v,w\rangle$ denotes the current position of the walker. Then $|v,w\rangle$ denotes a directed line from $v$ to $w$. At the same time, $|v,c\rangle$ also denotes a directed line from $v$, the target point of this directed line is based on the partition of $\overrightarrow{L}^{d}G$. Therefore, $|v,w\rangle$ corresponds to $|v,c\rangle$. And this fact builds a bridge between Szegedy QWM and coined QWM.

3. A coined QWM with the partition $\pi'$ and coin shift function $gc$ can be transformed to a Szegedy QWM.  $|v,c_j\rangle$ denotes a directed line from $v$ to $w$, where $(v,w)\in E(C'_j)$. Eq.\ref{Eq305} transform $|v,c_j\rangle$ to $|f'_{C'_j}(v),gc(v,c_j)\rangle$. Depend on Def.5, $f'_{C'_j}(v)=w$. $|f'_{C'_j}(v),gc(v,c_j)\rangle$ denotes a directed line from $w$ to $u$, where $(w,u)\in E(C'_k)$, $c_k=gc(v,c_j)$. Let $f(v,w)=(w,u)$, then  $f$ and the corresponding edge partition $\pi$  for a Szegedy QWM is fixed. The operator $S$ in Eq.\ref{Eq204} and $D$ in Eq.\ref{Eq305} have the same effect: $|v,w\rangle\longrightarrow |w,u\rangle$.  The requirement of $gc$ in Eq.\ref{Eq306} is to make sure the operator $D$ is unitary. Therefore, the edge partition $\pi$, fixed by $\pi'$ and $gc$, is an edge dicycle partition.

On the other hand,  a Szegedy QWM with an edge dicycle partition $\pi$ can be transformed to a coined QWM with any  partition $\pi'$.  The operator $S$ in Eq.\ref{Eq204} has the effect $|v,w\rangle\longrightarrow |w,u\rangle$. Let $(v,w)\in E(C'_j)$, then $|v,c_j\rangle$ denotes a same directed line from $v$ to $w$. hence, $f'_{C'_j}(v)=w$. Let $(w,u)\in E(C'_k)$, $gc(v,c_j)=c_k$. Then, the correspondence coined QWM with  the partition $\pi'$ and coin shift function $gc$ is fixed.

4. The operator $R$ in Eq.\ref{Eq203} has the following effect:{\setlength\arraycolsep{2pt}
\begin{eqnarray} \label{Eq310}
     && R: |v,w\rangle \rightarrow  \nonumber\\
     &&|v\rangle\otimes [(2q_{vw}-1)|w\rangle+\sum_{w'\neq w}2\sqrt{q_{vw}q_{vw'}} |w'\rangle         ]
\hspace{1mm}
\end{eqnarray}}With the relation of $|v,w\rangle$ and $|v,c\rangle$, $R$ can be transformed to {\setlength\arraycolsep{2pt}
\begin{eqnarray} \label{Eq311}
      R: |v,c\rangle \longrightarrow \sum_j A_{c,c_j} |v,c_j\rangle.
\hspace{1mm}
\end{eqnarray}}$A$ is in the form of
{\setlength\arraycolsep{2pt}
\begin{eqnarray} \label{Eq312}
      A=&&\left[
                \begin{array}{ccc}
                  2q_{vw}-1 & 2\sqrt{q_{vw}q_{vw'}}  \\
                  2\sqrt{q_{vw}q_{vw'}} & 2q_{vw'}-1\\
                \end{array}
      \right].
\hspace{1mm}
\end{eqnarray}} or
{\setlength\arraycolsep{2pt}
\begin{eqnarray} \label{Eq313}
      A=&&\left[
                \begin{array}{ccc}
                  2q_{vw'}-1 & 2\sqrt{q_{vw}q_{vw'}}  \\
                  2\sqrt{q_{vw}q_{vw'}} & 2q_{vw}-1\\
                \end{array}
      \right].
\hspace{1mm}
\end{eqnarray}}for two partitions, which are different at the position $v$, respectively. Therefore, by choosing proper coin operator, $R$ corresponds to $C$ in Eq.\ref{Eq304} with $A$ is a real operator. By extending $|\psi_v\rangle=|v\rangle\otimes \sum_w\sqrt{q_{vw}}|w\rangle$ in Eq.\ref{Eq205} to $|\psi_v\rangle=|v\rangle\otimes \sum_w \alpha_{vw}|w\rangle$, the coin operator $A$ in Ref.\ref{Eq312} can be extend to
{\setlength\arraycolsep{2pt}
\begin{eqnarray} \label{Eq314}
      A=&&\left[
                \begin{array}{ccc}
                  2|\alpha_{vw}|^2-1 & 2\alpha_{vw}^*\alpha_{vw'}  \\
                  2\alpha_{vw'}^*\alpha_{vw} & 2|\alpha_{vw'}|^2-1\\
                \end{array}
      \right]
\hspace{1mm}
\end{eqnarray}}which is a complex operator.

5. Till now, the way of relating the evolution operators of Szegedy QWM and coined QWM have been uncovered. The choice of initial state is also an element which affects the probability distribution. According to the relation of  $|v,w\rangle$ and $|v,c\rangle$, given the initial state for Szegedy QWM, the initial state for coined QWM can be constructed to produce the same probability distribution. It is vice versa.

In conclusion,  there are correspondence between Szegedy QWM and coined QWM. Furthermore, transforming the evolution of coined QWM to Szegedy QWM can help us to analyze coined QWM.

\section{Analysis of coined QWM}
\label{sec:level4}

In Ref. \cite{205}, coined QWM is controlled by a coin operator, a partition $\pi'$ and a proper coin shift function $gc$. Different partition will lead to different choice of  coin shift function and different QWM. However, by transforming coined QWM to Szegedy QWM, we find different QWM with different partition may be the same QWM in essence. With this fact, we should take a new look at coined QWM from another angle.

\subsection{ Essence of coined QWM}
\label{sec:level41}

Sect.\ref{sec:level32} shows the relation of coined QWM and Szegedy QWM. A coined QWM can be transformed to a Szegedy QWM, while a Szegedy QWM can be transformed to a coined QWM with any partition. That means all coined QWM can be transformed to a coined QWM with a same partition $\pi'$.

Partition $\pi2'$ in Fig.2 is a dicycle partition, which has some good properties for analysis. Therefore, we suggest researchers choose the partition $\pi2'$. Then, all coined QWM can be transformed to QWM with the dicycle partition $\pi2'$ and a coin shift function with the constraint:
{\setlength\arraycolsep{2pt}
\begin{eqnarray} \label{Euq306}
\{\overbrace{gc(v_i,c_k),\cdots}^m\}=\{c_1,c_2,\cdots,c_m\}\  with \ v=f'_{C_k'}(v_i).\
\hspace{1mm}
\end{eqnarray}}

Then the elements which affect the evolution of coined QWM is the coin operator and the coin shift function $gc$. At the same time, the evolution of Szegedy QWM is  affected by the adjacent matrix  and the edge dicycle partition. Each form of QWM have its advantages. Coined QWM has concise form for experiment and designing algorithm. Szegedy QWM shows the essential evolution of QWM.  Furthermore, except coined QWM with partition $\pi2'$, other form of coined QWM have their special meanings. The QWM in Ref.\cite{201}  was presented by considering  recycled coins.  The QWM in Ref.\cite{202,203}  was presented by using the coin state to decides the shift is `Reflect' or `Transmit'. Our results does not mean that other form of coined QWM do not have value to study.

\subsection{QWM1 and QMW2}
\label{sec:level42}

There are two kinds of coined QWM on the line in Ref.\cite{201} and Ref.\cite{203,204}, respectively. The two coined QWM were the only two QWM before the generic model of coined QWM was presented in Ref.\cite{205}. However, through the analysis in Sect.\ref{sec:level32}, these two quantum walks with 1 memory are in fact the same one when they have a proper position-dependent coin operator.

QWM1 is a coined QW with 1 memory  with partition $\pi1'$ in Fig.1 and $gc_1$, which is  as follows. \begin{eqnarray} \label{Euq321}
   &&gc_1(v_{x,x+1},1)=1, \ \ \ gc_1(v_{x,x-1},1)=-1, \nonumber\\
   &&gc_1(v_{x,x+1},-1)=1, \ gc_1(v_{x,x-1},-1)=-1,
\hspace{1mm}
\end{eqnarray}\begin{figure*}[!ht]\label{Fig601}
 \begin{center}
  \subfigure[]{
\label{Fig001a}
\includegraphics[width=8cm]{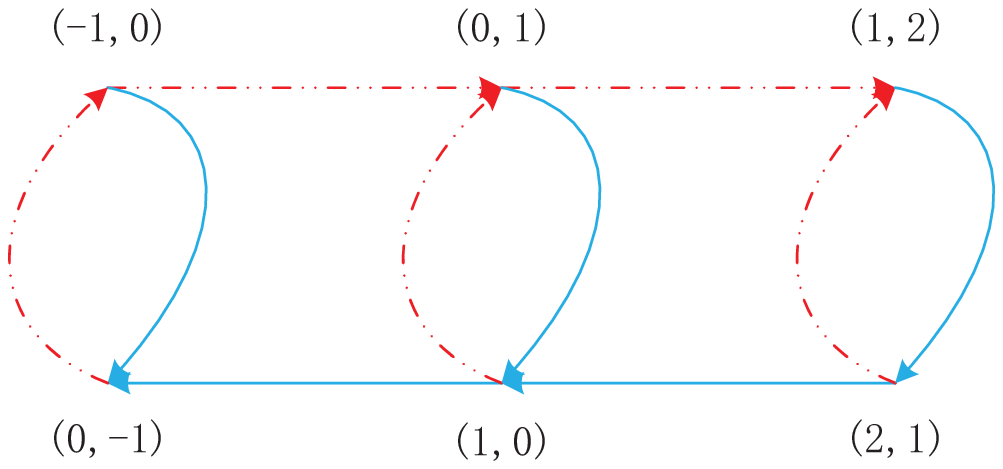}}
  \subfigure[]{
  \label{Fig001b}
 \includegraphics[width=8cm]{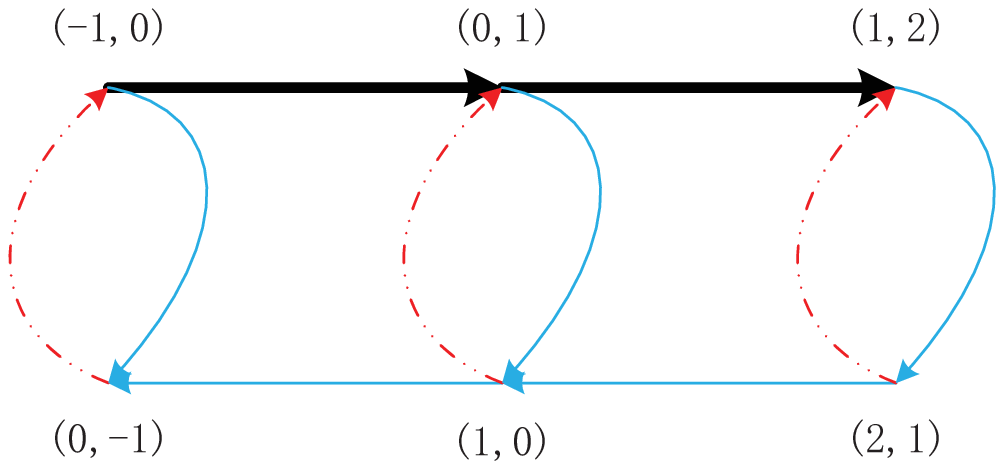}}
  \subfigure[]{
  \label{Fig001c}
\includegraphics[width=8cm]{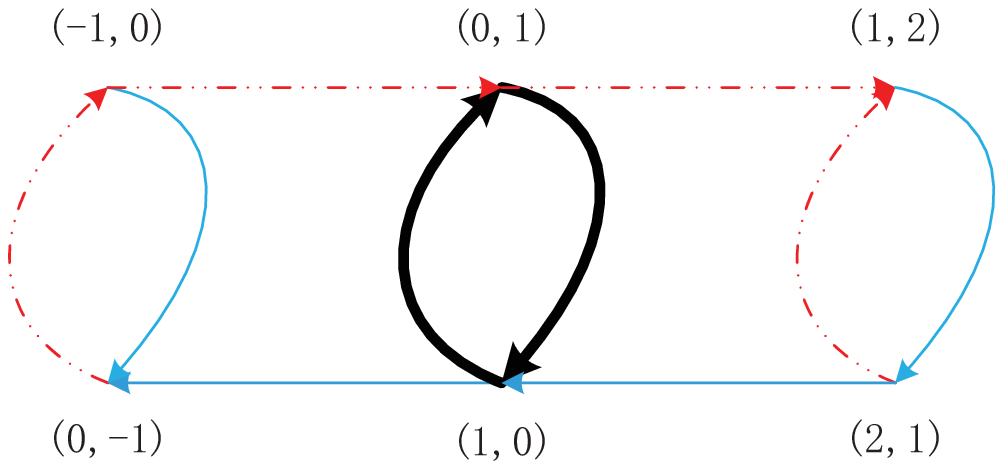}}
  \subfigure[]{
  \label{Fig001c}
\includegraphics[width=8cm]{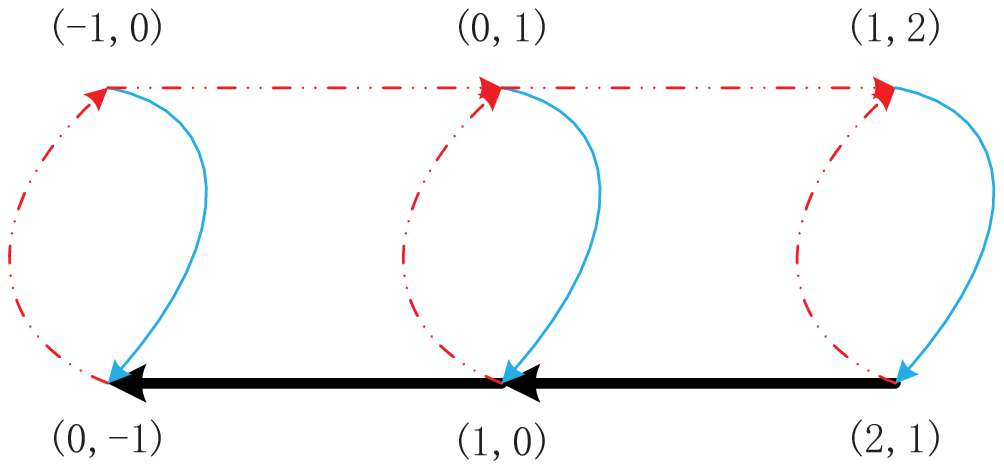}}
  \end{center}
  \renewcommand{\figurename}{Fig.}
  \caption{Partition $\pi1'$ of the line digraph. Black lines show the movement of QWM1.}
\end{figure*}
Let the coin operator be\begin{equation}\label{Euq322}
              H1=\left\{
                             \begin{array}{ccc}
                \left[
                \begin{array}{ccc}
                  1/\sqrt2 & 1/\sqrt2  \\
                  1/\sqrt2 & -1/\sqrt2\\
                \end{array}
               \right]                     &  at\ positions\  $(x,x+1)$; \\
                \left[
                \begin{array}{ccc}
                  -1/\sqrt2 & 1/\sqrt2  \\
                  1/\sqrt2 & 1/\sqrt2\\
                \end{array}
                 \right]                  &  at\  positions\  $(x+1,x)$.  \\
                              \end{array}
                \right.
\end{equation}

QWM2 is a coined QW with 1 memory with  a dicycle partition $\pi2'$ in Fig. 2 and $gc_2 (v,c_j)=c_j$. Let the coin operator be{\setlength\arraycolsep{2pt}
\begin{eqnarray} \label{Euq323}
      H2=&&\left[
                \begin{array}{ccc}
                  1/\sqrt2 & 1/\sqrt2  \\
                  1/\sqrt2 & -1/\sqrt2\\
                \end{array}
      \right].
\hspace{1mm}
\end{eqnarray}}

\begin{figure*}[!ht]\label{Fig602}
 \begin{center}
  \subfigure[]{
\label{Fig001a}
\includegraphics[width=8cm]{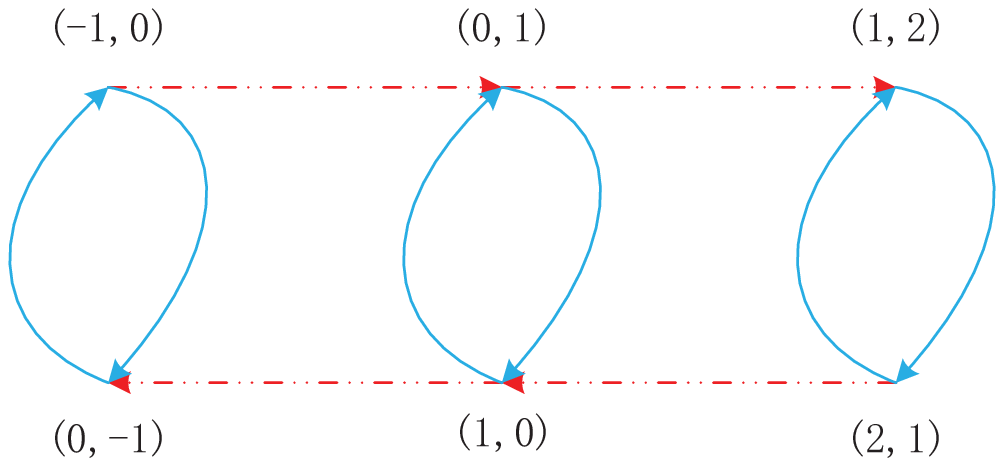}}
  \subfigure[]{
  \label{Fig001b}
 \includegraphics[width=8cm]{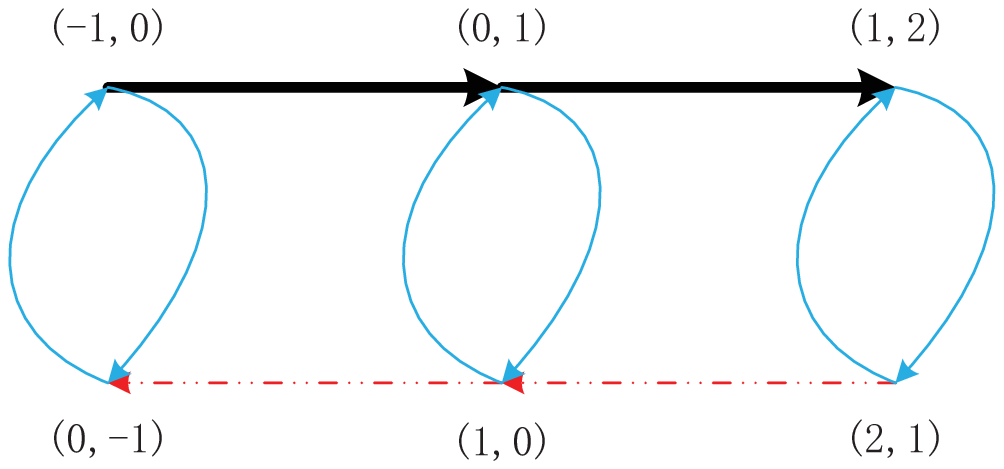}}
  \subfigure[]{
  \label{Fig001c}
\includegraphics[width=8cm]{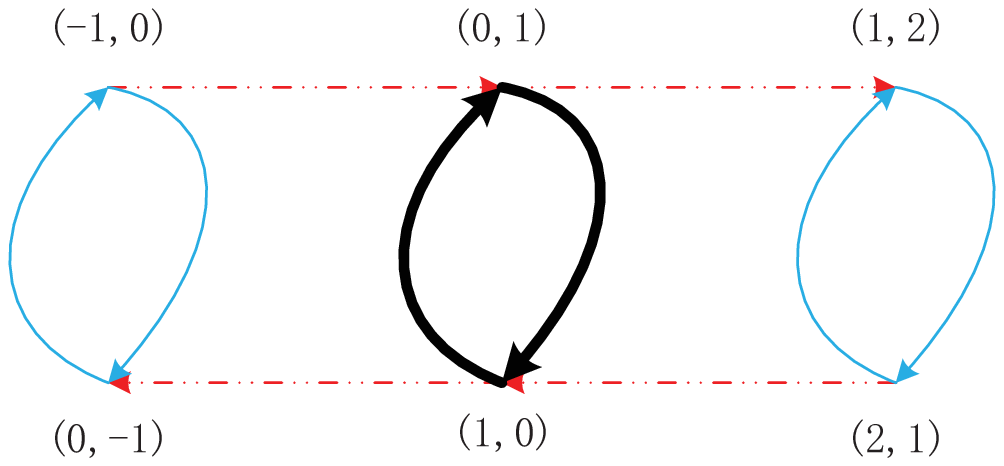}}
  \subfigure[]{
  \label{Fig001c}
\includegraphics[width=8cm]{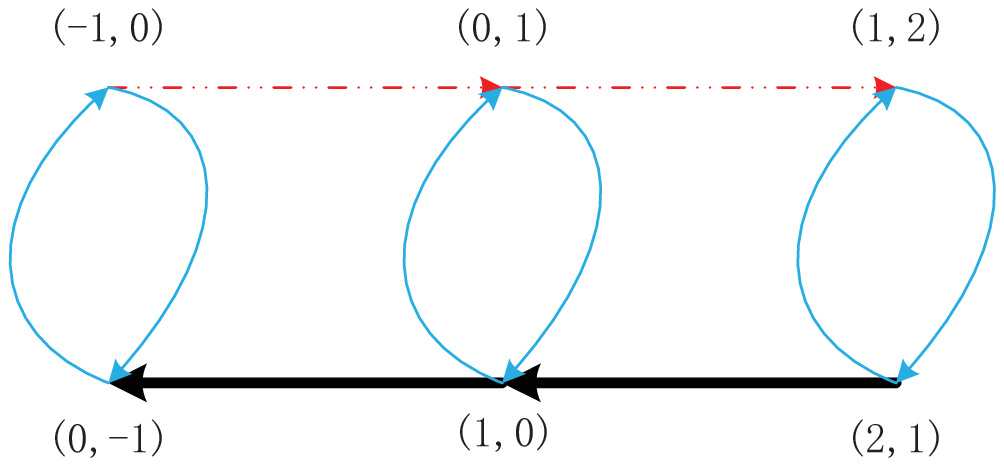}}
  \end{center}
  \renewcommand{\figurename}{Fig.}
  \caption{Partition $\pi2'$ of the line digraph. Black lines show the movement of QWM2.}
\end{figure*}

In Fig.1(a), the partition $\pi_1'$ of $\overrightarrow{L}G$ is shown. According to the operator $D$ in Eq.\ref{Eq305}, three kinds of movement of QWM1 are shown by black arrows in Fig.1(b,c,d), while three kinds of movement of QWM2 are shown by black arrows in Fig.2(b,c,d). The movements of QWM1 and QWM2 are exactly the same. However, $\pi_1'$ and $\pi_2'$ are different at lower positions of $\overrightarrow{L}G$, such as positions $(0,-1)$, $(1,0)$, $(2,1)$. Therefore, by choosing two different operators at these positions in the form of  Eq.\ref{Eq312} and Eq.\ref{Eq313}, QWM1 and QWM2 have the same evolution.

Then,  the initial states $\alpha|x-1,x,1\rangle+\beta|x-1,x,-1\rangle+\alpha'|x+1,x,1\rangle+\beta'|x+1,x,-1\rangle$ for QWM1 and $\alpha|x-1,x,1\rangle+\beta|x-1,x,-1\rangle+\beta'|x+1,x,1\rangle+\alpha'|x+1,x,-1\rangle$ can generate same probability distribution. It is also backed up by  simulation results.

\section{Summary}
\label{sec:level5}

QWM are a type of modified quantum walks that have many extra coins to record the walker's latest path. In this paper, we present the general model of Szegedy QWM. And the relation of Szegedy QWM and coined QWM has been uncovered. Furthermore, transforming coined QWM to Szegedy QWM provides a new angle for analyzing coined QWM.

Through the analysis, we expose the fact that a coined QWM can be transformed to a Szegedy QWM, while a Szegedy QWM can be transformed a coined QWM with any partition. That means all coined QWM can be transformed to a coined QWM with a same partition. Furthermore, two different coined QWM, which attract much attention,  have same evolution when they have a position-dependent coin operator.

\begin{acknowledgments}
This work is supported by NSFC (Grant Nos. 61701229, 61571226, 61572053, 61702367), Natural Science Foundation of Jiangsu Province, China (Grant No. BK20170802), China Postdoctoral Science Foundation funded Project (Grant Nos. 2018M630557, 2018T110499), Jiangsu Planned Projects for Postdoctoral Research Funds (Grant No. 1701139B), the Beijing Natural Science Foundation (Grant No. 4162005), The Research Project of Tianjin Municipal Commission of Education(Grant No. 2017KJ033),  the Fundamental Research Funds for the Central Universities (Grant No. 2018RC55),  the Beijing Talents Foundation (Grant No. 2017000020124G062).

\end{acknowledgments}

\bibliography{basename of .bib file}
\bibliography{apssamp}

\begin{thebibliography}{99}

\bibitem{101} A. Ambainis, E. Bach, A. Nayak, A. Vishwanath, and J. Watrous,  STOC '01 Proceedings of the thirty-third annual ACM symposium on Theory of computing (ACM New York, NY, USA) pp. 37-49 (2011).

\bibitem{102} A. Nayak, and A. Vishwanath,  arXiv: quant-ph/0010117 (2000).

\bibitem{103} C. I. Chou, and C. L. Ho,  Chin. Phys. B \textbf{23}, 110302 (2014).

\bibitem{104} M. Li, Y. S. Zhang, and G. C. Guo, Chin. Phys. B \textbf{22}, 030310 (2013).

\bibitem{105} P. Xue, and B. C. Sanders,  Phys. Rev. A \textbf{85}, 022307 (2012).

\bibitem{106} C. Di Franco, M. Mc Gettrick, and T. Busch, Phys. Rev. L \textbf{106}, 080502 (2011).

\bibitem{107} C. Di Franco, M. Mc Gettrick, T. Machida, and T. Busch,  Phys. Rev. A \textbf{84}, 042337 (2011).

\bibitem{108} N. Inui, N. Konno, and E. Segawa,  arXiv: quant-ph/0507207v1 (2005).

\bibitem{109} D. Li, M. Mc Gettrick, W.W. Zhang, K.J. Zhang, Chin. Phys. B, \textbf{24}, 050305 (2015).

\bibitem{110} D. Li, J. Zhang, F. Z. Guo, W. Huang, Q. Y. Wen, and H. Chen, Quant. Inf. Proc. \textbf{12}, 1501-1513 (2013).

\bibitem{111} D. Li, Y.G. Yang, J.L. Bi, J.B. Yuan, J. Xu, Scientific Reports \textbf{8}, 225 (2018).

\bibitem{112} P. P. Rohde, A. Schreiber, M. Stefanak, I. Jex, and C. Silberhorn, New J. Phys. \textbf{13}, 013001 (2011).

\bibitem{113} K. Mayer, M. C. Tichy, F. Mintert, T. Konrad, and A. Buchleitner, Phys. Rev. A \textbf{83}, 062307 (2011).

\bibitem{114} R. Zhang, H. Qin, B. Tang, and P. Xue, Chin. Phys. B \textbf{22}, 110312 (2013).

\bibitem{115} R. Zhang, Y. Q. Xu, and P. Xue,  Chin. Phys. B \textbf{24}, 010303 (2015).



\bibitem{117} A. Ambainis, arXiv: quant-ph/0311001 (2003).

\bibitem{118} N. Shenvi, J. Kempe, and K. Birgitta Whaley, Phys. Rev. A \textbf{67}, 052307 (2003).

\bibitem{119} B. Hein, and G. Tanner, Phys. Rev. A   \textbf{82}, 012326 (2010).

\bibitem{120} S. D. Berry and J. B. Wang, Phys. Rev. A \textbf{82}, 042333 (2010).

\bibitem{121} L. Tarrataca,  and A. Wichert,  Quant. Inf. Proc. \textbf{12}, 1365-1378 (2013).

\bibitem{122} S. D. Berry, and J. B. Wang, Phys. Rev. A \textbf{83}, 042317 (2011).

\bibitem{123} B. L. Douglas, and J. B. Wang, J. Phys. A \textbf{41}, 075303 (2008).



\bibitem{201} P. P. Rohde, G. K. Brennen, and A. Gilchrist, Phys. Rev. A, \textbf{87}, 052302 (2013).

\bibitem{202} M. Mc Gettrick, Quant. Inf. Compu.,  \textbf{10}, 0509-0524 (2010).

\bibitem{203} M. Mc Gettrick, J. A. Miszczak, Physica A, \textbf{399},  163-170 (2014).

\bibitem{204} N. Konno, T. Machida, Quant. Inf. Compu., \textbf{10}, 1004-1017 (2010).


\bibitem{205} D. Li, M. Mc Gettrick, F. Gao, J. Xu and Q.Y. Wen, Phys. Rev. A \textbf{93}, 042323 (2016).










\bibitem{301} M. Szegedy, FOCS '04 Proceedings of the  45th Annual IEEE Symposium on Foundations of Computer Science, 32-41 (2004).


\bibitem{302} R. Portugal, Quant. Inf. Proc. \textbf{15}, 1387-1409 (2016).

 \bibitem{303} N. Konno, R. Portugal, I. Sato, E. Segawa,  Quant. Inf. Proc. 17:100 (2018).











\bibitem{207} Y. Liu, J.B. Yuan, B.J. Duan and D. Li,  scientific reports. \textbf{8}, 9548 (2018).





\bibitem{401}	H. Buhrman, R. Spalek, Proceeding SODA '06 Proceedings of the seventeenth annual ACM-SIAM symposium on Discrete algorithm.  880 (2006).

\bibitem{402}	F. Magniez, M. Santha, M. Szegedy, SIAM J. Comput. 37(2), 413¨C424 (2007).

\bibitem{403}	F. Magniez,  A. Nayak, Algorithmica 48:221 (2007).

\bibitem{404}	G. Wang, Quantum Information and Computation 17 (11 and 12) 987-1026 (2017).


\bibitem{405}	G.D. Paparo, M. Muller, F. Comellas, M. Angel, M.Delgado,  Scientific Reports \textbf{3}, 2773 (2013).









\end{thebibliography}

\end{document}